\documentclass{aa}
\usepackage{graphicx}
\usepackage{hyperref}
\usepackage{natbib}
\usepackage{float}
\usepackage{longtable}
\usepackage{csvsimple}
\usepackage{listings}
\usepackage{color}
\usepackage{textcomp}
\usepackage{caption}
\usepackage{subfig}
\usepackage[retainorgcmds]{IEEEtrantools}
\usepackage[utf8]{inputenc}
\usepackage[varg]{txfonts}

\definecolor{dkgreen}{rgb}{0,0.6,0}
\definecolor{gray}{rgb}{0.5,0.5,0.5}
\definecolor{mauve}{rgb}{0.58,0,0.82}

\usepackage{hyperref}
\hypersetup{
    colorlinks = true,
    linkcolor = {blue}, 
    citecolor = {blue},
    urlcolor = {blue}
}

\begin{document}
    \title{Near real-time precipitable water vapour monitoring for correcting near-infrared observations using satellite remote sensing}
    
    \titlerunning{Satellite imagery near real-time PWV correction}
    \author{E.A. Meier Vald\'es
        \inst{1}
        \and 
        B.M.~Morris
        \inst{1}
        \and 
        B.-O. Demory
        \inst{1}}
    \institute{Center for Space and Habitability (CSH), University of Bern,
              Gesellschaftsstrasse 6, 3012 Bern, Switzerland\\
              \email{erik.meiervaldes@csh.unibe.ch}
              }
\date{March 2021}

\abstract
{In the search for small exoplanets orbiting cool stars whose spectral energy distributions peak in the near infrared, the strong absorption of radiation in this region due to water vapour in the atmosphere is a particularly adverse effect for the ground-based observations of cool stars.}
{To achieve the photometric precision required to detect exoplanets in the near infrared, it is necessary to mitigate the impact of variable precipitable water vapour (PWV) on radial-velocity and photometric measurements. The aim is to enable global PWV correction by monitoring the amount of precipitable water vapour at zenith and along the line of sight of any visible target.}
{We developed an open source Python package that uses Geostationary Operational Environmental Satellites (GOES) imagery data, which provides temperature and relative humidity at different pressure levels to compute near real-time PWV above any ground-based observatory covered by GOES every 5 minutes or 10 minutes depending on the location.}
{We computed PWV values on selected days above Cerro Paranal (Chile) and San Pedro M\'artir (Mexico) to benchmark the procedure. We also simulated different pointing at test targets as observed from the sites to compute the PWV along the line of sight. To asses the accuracy of our method, we compared our results with the on-site radiometer measurements obtained from Cerro Paranal.}
{Our results show that our publicly-available code proves to be a good supporting tool for measuring the local PWV for any ground-based facility within the GOES coverage, which will help in reducing correlated noise contributions in near-infrared ground-based observations that do not benefit from on-site PWV measurements.}


\maketitle

\section{Introduction}
\label{section:introduction}

In our search for potential habitable exoplanets, cool stars \--- such as the TRAPPIST-1 system \citep{gillon,Luger:2017} \--- have proven to be promising candidates. In the past years, numerous detections of exoplanets have been reported and projects such as SPECULOOS \citep{Jehin:2018,Delrez:2018} aim to detect small exoplanets orbiting ultra-cool stars for which the spectral energy distribution peak in the near infrared. A key requirement to detect small exoplanets orbiting cool stars is to efficiently correct the adverse effects induced by the Earth's atmosphere. Water vapour is one of the main atmospheric gases that absorbs incoming radiation in the near infrared (NIR) region, specifically around 1 $\mu$m and 5 $\mu$m \citep[see][chap. 6]{peixoto}. This characteristic of water vapour poses a serious challenge for ground-based observations in the NIR region. The amount of precipitable water vapour is variable in space and time, thus inducing correlated noise in the photometric time-series, potentially masking or mimicking an exoplanet transit \citep{baker}. In the search for exoplanets via the radial velocity (RV) method, telluric lines can limit the measurement accuracy at the level of 1 m/s \citep{cunha}, which cannot be neglected for instruments aiming to achieve sub m/s precision, such as ESPRESSO \citep{Pepe:2020}. The above-mentioned issues justify the interest in monitoring the temporal variability of precipitable water vapour (PWV) during observations to better understand and correct correlated noise due to the atmosphere in the search for small planets.

Several studies (\citet{kassomenos, otarola, buehler, marin}) have focused on monitoring and characterising precipitable water vapour variability from a meteorological point of view, for which daily, monthly or seasonal values are of interest. Thus, the temporal resolution between measurements may be lower (e.g. retrieval of data every 12 or 24 hours) than those studies with an astronomical focus. While average PWV above certain locations is an important factor for astronomical site selection, much higher temporal resolution of the order of minutes is required to study PWV variability over timescales relevant for RV or photometric exoplanet observations. In particular, the relevant timescales for photometric observations is the transit duration, which has been observed to last less than an hour for some exoplanets \citep[see][]{Gillon:2017}. A lot of effort has been put into mitigating the effect of water vapour absorption on astronomical observations. \citet{baker} developed an automatic PWV monitoring multiband photometer system and installed it at the Fred Lawrence Whipple Observatory in Arizona. The instrumentation achieves better than 0.5 mm precision for PWV conditions below 4 mm and the data was compared to GPS monitoring systems. A GPS is commonly used to measure PWV by means of the delay of satellite signals which can be formulated as an excess length and usually determined at the zenith above the ground-based receiver station. This parameter, called Zenith Total Delay (ZTD) includes a wet delay component, which is a function of the water vapour distribution \citep{mile}. The issue of measuring PWV from GPS data arises on the uncertainty in the ZTD measurement, since the standard deviation for PWV is approximately the same for all measurements. Therefore, the relative uncertainty is high for low values of PWV, but low for high values of PWV \citep{buehler}. The GPS method is therefore less accurate in dry conditions, which are typical of most astronomical observing sites. In this paper we define a region as dry for an average of PWV below 4 mm. 

Additionally, \citet{buehler} conducted a multi-instrument comparison of PWV over the subarctic site Kiruna (Sweden). The different sensors and retrieval methods used involved radiosondes, GPS, ground-based Fourier-transform infrared spectrometer, ground-based microwave radiometer and satellite-based microwave radiometer. The available parameters for each measurement taken by the radiosonde include time of flight, pressure, temperature, relative humidity, height and dew point, but measurements are limited to periods of scheduled launches. To estimate PWV using Fourier Transform Infrared spectroscopy, solar absorption spectra are recorded and processed via radiative transfer models, but this technique requires cloud-free conditions and measurements are limited to periods when the sun is above the horizon. The ground-based radiometer measures thermal emission spectra to monitor stratospheric trace gases. To retrieve ozone profiles the spectrum has to be corrected for varying tropospheric water vapour, resulting in an offset. Then models based on radiative transfer theory are used to calculate a water vapour profile matching the offset and actual temperature and pressure values. Finally, since the atmosphere is too dry in polar regions, the satellite-based radiometer measures brightness temperature -instead of relative humidity- and relates it with PWV. The results agree reasonably well, differing by values below $\pm$ 1 mm, except for ground-based measurements without filtering out cloud presence \citep{buehler}. Additionally, given that the systematic differences depend strongly on the instrument, location and method, it is not conclusive which technique is most suitable for estimating PWV. Additionally, which technique is most suitable from a scientific point of view depends on practical factors such as maintenance costs.

Andre Erasmus conducted several studies on forecasting of precipitable water vapour in Chile. \citet{erasmus1} did a survey using 58 months of meteorological satellite data taken every 12 hours between July 1993 and September 1999 in Northern Chile to evaluate the best site for the Extremely Large Telescope (ELT) based on cloud cover and water vapour. This work concluded that Paranal, Quimal, Yacas and Cascasco were the best candidates based on cloud cover. Cerro Chasc\'on was found to be the best site based on PWV analysis. The ELT is now under construction on Cerro Armazones. Researchers have also made use of older GOES imagery to compute PWV values every 6 hours \citep{erasmus2} or every 3 hours \citep{marin} by means of radiance data. Planck functions relate radiance with brightness temperature and the method to compute PWV involves a semi-empirical formula valid only for the 6.7 $\mu$m channel between 600 and 300 hPa pressure levels. We note that this approach is no longer valid using data of present operational GOES satellites, since they are not equipped with a 6.7 $\mu$m channel. Instead, they take measurements at bands with central wavelength of 7.34, 6.93 and 6.17 $\mu$m. This allows to measure low- (from sea-level to approximately 900 hPa), mid- (from approximately 900 to 700 hPa) and high-level (from 700 to 300 hPa) water vapour, respectively \citep{PUG5}.

\citet{querel} used the Low Humidity and Temperature Profiling (LHATPRO) microwave radiometer, installed at ESO Cerro Paranal observatory in support of the Very Large Telescope (VLT), to monitor the amount of PWV above the site down to an elevation angle of 27.5 degrees every 6 hours during a period of 21 months. They found that PWV across the sky over Paranal is very uniform, suggesting that constraining PWV measurements at the zenith would be sufficient to support most science applications. Another important result is that the highest spatial and temporal variation of PWV occurs for low PWV conditions. Finally they recommend providing simultaneous line of sight observations pointed towards the target.

A Python package that estimates PWV was developed by \citet{perrefort}, which provides models for atmospheric transmission due to water vapour at Kitt Peak National Observatory, Arizona, or user specified sites in the continental United States region and most of central America affiliated to the SuomiNet project \footnote{\url{https://www.suominet.ucar.edu/}}. Given a date and airmass, the package determines PWV column density using data regarding the delay in GPS signals published every hour.

There is no defined boundary set to compute PWV. While the total PWV value is computed between sea-level atmospheric pressure and approximately 100 hPa, this is not the ideal choice when estimating PWV on high-altitude sites, since it does not take into account the actual pressure on the surface, given that pressure decreases with height. Applying this definition at Cerro Paranal, where the mean surface pressure is 750 hPa, would lead to an overestimated PWV value due to contribution underneath this pressure level. \citet{erasmus3} set an upper limit of 300 hPa because at lower pressure the contribution from PWV is very small (see Sect. \ref{section:PWV above 300 hPa}). \citet{marin} estimated the integrated PWV in the Chajnantor plateau (5100 m above sea level) in Chile for pressure levels from 550 hPa to 100 hPa, since the mean surface pressures at the study sites ranged between 550 hPa and 500 hPa over the year and the upper pressure level corresponds to the highest one available in the dataset they used.

The aim of the present study is to present \texttt{fyodor} \footnote{\texttt{fyodor} can be downloaded at \url{https://github.com/Erikmeier18/fyodor}}, an open-source Python package that uses GOES imagery data to compute near real-time PWV on any site covered by the GOES-W/E (Americas). The main motivation is to assist data analyses of observations of cool stars in the near infrared region, including the SAINT-EX Observatory \citep{Demory:2020} located in San Pedro M\'artir (Baja California, Mexico). The novelty of our approach is that on the user's request, the code computes the amount of precipitable water vapour at the zenith or along the line of sight to a target in the sky. To asses the program's performance, ESO's Cerro Paranal (Chile) Observatory was chosen to compare the results with the measurements obtained by the LHATPRO instrument at the same location.

\section{Methodology}
\label{section:method}

We present the instrumentation and methodology in this section. First, in Sect. \ref{section:goes} and \ref{section:products} we introduce the currently operational satellites of the National Oceanic and Atmospheric Administration (NOAA) and the relevant data products with its range, accuracy and performance. Then, Sect. \ref{section:abi} is dedicated to the coordinate system used by GOES satellites. In Sect. \ref{section:water vapour} we present the theoretical ground on precipitable water vapour, along with the complete derivation in the Appendix Sect. \ref{section:derivation}. Finally, the proposed method on computing PWV along the line of sight to a target in the sky is explained in Sect. \ref{section:Precipitable water vapour along the line of sight to the target}.  

\subsection{GOES-R Series}
\label{section:goes}

NOAA currently operates two Geostationary Operational Environmental Satellites (GOES), at 75 degrees west and 137 degrees west longitude, to provide continuous weather imagery and monitoring of meteorological data across the U.S.A. and most part of the American continent. The current generation is called GOES-R Series and the primary instrument is the Advanced Baseline Imager (ABI), a multi-spectral channel, two-axis scanning radiometer that provides geolocated observations on three standard coverage regions: \textit{Full Disk} defined as a circle with 17.4 degree angular diameter from the perspective of the satellite covering a near hemispheric Earth region; \textit{CONUS} covers an area of 10N-60N latitude and 60W-125W longitude and \textit{Mesoscale}, equivalent to a region of approximately 1000x1000 kilometres. ABI Bands 1 to 6 measure solar reflected radiance at visible and near infrared wavelengths, while bands 7 to 16 measure emitted radiance at infrared wavelengths. All data are made available in the netCDF-4 file format \citep{PUG1}. For this study we will make use of data obtained by the GOES-16 satellite.  

\subsection{GOES-R data products}
\label{section:products}

Among all data products provided by NOAA, which are publicly available to download\footnote{\url{https://www.avl.class.noaa.gov/saa/products/search?datatype_family=GRABIPRD}}, we require only temperature, pressure and relative humidity data to compute the PWV. The data products used are labelled as Legacy Vertical Temperature Profile (LVTP) and Legacy Vertical Moisture Profile (LVMP), which make use of all infrared channels (Bands 7-16). Both data products contain a three-dimensional variable with pixel values identifying the air temperature and relative humidity at 101 discrete standard pressure levels. Each file corresponds to one measurement for all pressure levels and the measurements are taken every 10 minutes for the Full Disk region and every 5 minutes for the continental United States (CONUS) region. Temperature values are provided in Kelvin, while relative humidity is given as a fraction ranging between 0 and 1. The spatial resolution of both data products is 10 km and data are generated if the following criteria are met:

\begin{itemize}
    \item Clear sky
    \item Geolocated source data to local zenith angles of 80 degrees for both daytime and nighttime conditions (angle between the line of sight to the satellite and the local zenith at the observation target).
\end{itemize}

The temperature range is between 180 and 320 K with an accuracy of 1 K and a precision of 2 K between the top of the boundary layer and 400 hPa. Regarding relative humidity both accuracy and precision are 0.18 from the surface up to 300 hPa and 0.2 between 300 and 100 hPa.  
    
Conditions for good quality data also require latitude $|\ell| \leq 70^\circ$ \citep{PUG5}. Before the data are released, internal algorithms check the data for its quality. If a pixel does not fulfil the conditions, the data point is masked. It is very important to note that of the temperature and moisture values taken at 101 pressure levels, only 54 temperature and 35 moisture pressure levels are actually measured. These temperature levels are from approximately 103 hPa to 1014 hPa, while the moisture levels are from 300 hPa to 1014 hPa. The rest of the levels are retrieved via interpolation.

Additionally, NOAA offers a product containing the integrated column water vapour amount from the surface to a height corresponding to a pressure of 300 hPa called Total Precipitable Water (hereafter TPW). It has an accuracy of 1 mm and precision of 3 mm. The advantages and disadvantages of using this end product are discussed in Sect. \ref{section:results}.

\subsection{ABI Fixed grid}
\label{section:abi}

The coordinate system used by the GOES-R, called Advanced Baseline Imager (ABI) Fixed grid, is a projection based on the viewing perspective of the location of the satellite in geostationary orbit. Data points at a particular horizontal spatial resolution on the fixed grid have the same angular separation from the satellite's perspective. Depending on the data product's resolution, Earth is covered by equally sized squares, each one of them populated by data. The x-axis represents the ABI East/West scan angle, y-axis represents North/South scan angle. GOES-16 products with a spatial resolution of 10 km at nadir have data points with an angular separation of 280 $\mu$radians. To locate a specific pixel in a file, one has to give its horizontal and vertical index. The dimensions of a data product with 10 km horizontal resolution is 1086x1086.

To navigate through data points in the ABI Fixed grid, one has to perform a transformation between latitude and longitude coordinates and ABI Fixed grid scan angle coordinates. The equations presented here assume points lying on the Geodetic Reference System 1980 ellipsoid. Given geodetic latitude $\phi$ and longitude $\lambda$ in radians, the scan angles x and y are computed by the following equations \citep{PUG5}:
\begin{IEEEeqnarray*}{rll}
\label{eqn:scanangles}
    &x& = \arcsin\left(\frac{-s_{y}}{\sqrt{s_{x}^{2}+s_{y}^{2}+s_{z}^{2}}}\right)  \\
    &y& = \arctan\left(\frac{s_{z}}{s_{x}}\right) \IEEEyesnumber
\end{IEEEeqnarray*}
where $s_{x}$ is the x-axis from the satellites' reference frame and is defined as the line from the satellite to the centre of the Earth, $s_{y}$ is the y-axis aligned with the equatorial axis and $s_{z}$ completes the orthogonal coordinate system and is parallel to the line that passes from the centre of the Earth to the north pole. These quantities are computed as follows:
\begin{IEEEeqnarray*}{rll}
\label{eqn:scanangles2}
    &s_{x}& = H-r_{c}\cos(\phi_{c})\cos(\lambda-\lambda_{0}) \\
    &s_{y}& = -r_{c}\cos(\phi_{c})\sin(\lambda-\lambda_{0}) \\
    &s_{z}& = r_{c}\sin(\phi_{c}) \IEEEyesnumber
\end{IEEEeqnarray*}
$H$ is the satellite's height from the centre of Earth in meters, $\lambda_{0}$ is the satellite's longitude, also called longitude of projection origin. These quantities, as well as Earth's semi major axis $r_{eq}$, semi minor axis $r_{pol}$ and eccentricity $e$ are included in every GOES netCDF-4 file. The geocentric latitude $\phi_{c}$ and geocentric distance to the point on the ellipsoid $r_{c}$ are
\begin{IEEEeqnarray*}{rll}
\label{eqn:scanangles3}
    &\phi_{c}& = \arctan\left(\frac{r_{pol}^{2}}{r_{eq}^{2}}\tan(\phi)\right) \\
    &r_{c}& = \frac{r_{pol}}{\sqrt{1-e^{2}\cos^{2}(\phi_{c})}}. \IEEEyesnumber
\end{IEEEeqnarray*}

\subsection{Definition of precipitable water vapour}
\label{section:water vapour}

Among the variable gases, water vapour is predominant in the lower troposphere, comprising up to 4\% of the volume of air in the atmosphere \citep{ahrens}. The amount of water vapour contained in a unit area column of air between two pressure levels is given by the expression \citep{kassomenos}
\begin{IEEEeqnarray}{rll}
\label{eqn:PWV2}
    PWV = -\frac{1}{g \rho_{w}}\int_{p_{i}}^{p_{i+1}} q ~ dp,
\end{IEEEeqnarray}
where $q$ is the specific humidity, $p$ is pressure and $g$ the acceleration due to Earth's gravity. It represents the amount of liquid water that would result if all the water vapour in the column were condensed and is usually expressed in units of kg m$^{-2}$ or mm after dividing the expression above by the density of water $\rho_{w}$.

A negative sign has to be included in front of the integral if we define that for ascending index $i=1,2,3, \ldots $, the absolute value of the pressure decreases. This convention will be adopted from now on since pressure decreases with increasing altitude.

The specific humidity can be expressed as a function of temperature, pressure and relative humidity as follows (see Appendix Sect. \ref{section:derivation} for the derivation and details):
\begin{IEEEeqnarray*}{rll}
\label{eqn:PWV3}
    PWV &=& -\frac{1}{g \rho_{w}}\int_{p_{i}}^{p_{i+1}} \frac{0.622u \cdot 6.112 \exp \left(\frac{17.67 T}{T+243.5}\right)}{p-0.378u\cdot 6.112 \exp \left(\frac{17.67 T}{T+243.5}\right)} dp. \IEEEyesnumber
\end{IEEEeqnarray*}

\subsection{Precipitable water vapour along the line of sight to the target}
\label{section:Precipitable water vapour along the line of sight to the target}

\begin{figure}
    \centering
    \resizebox{\hsize}{!}{\includegraphics{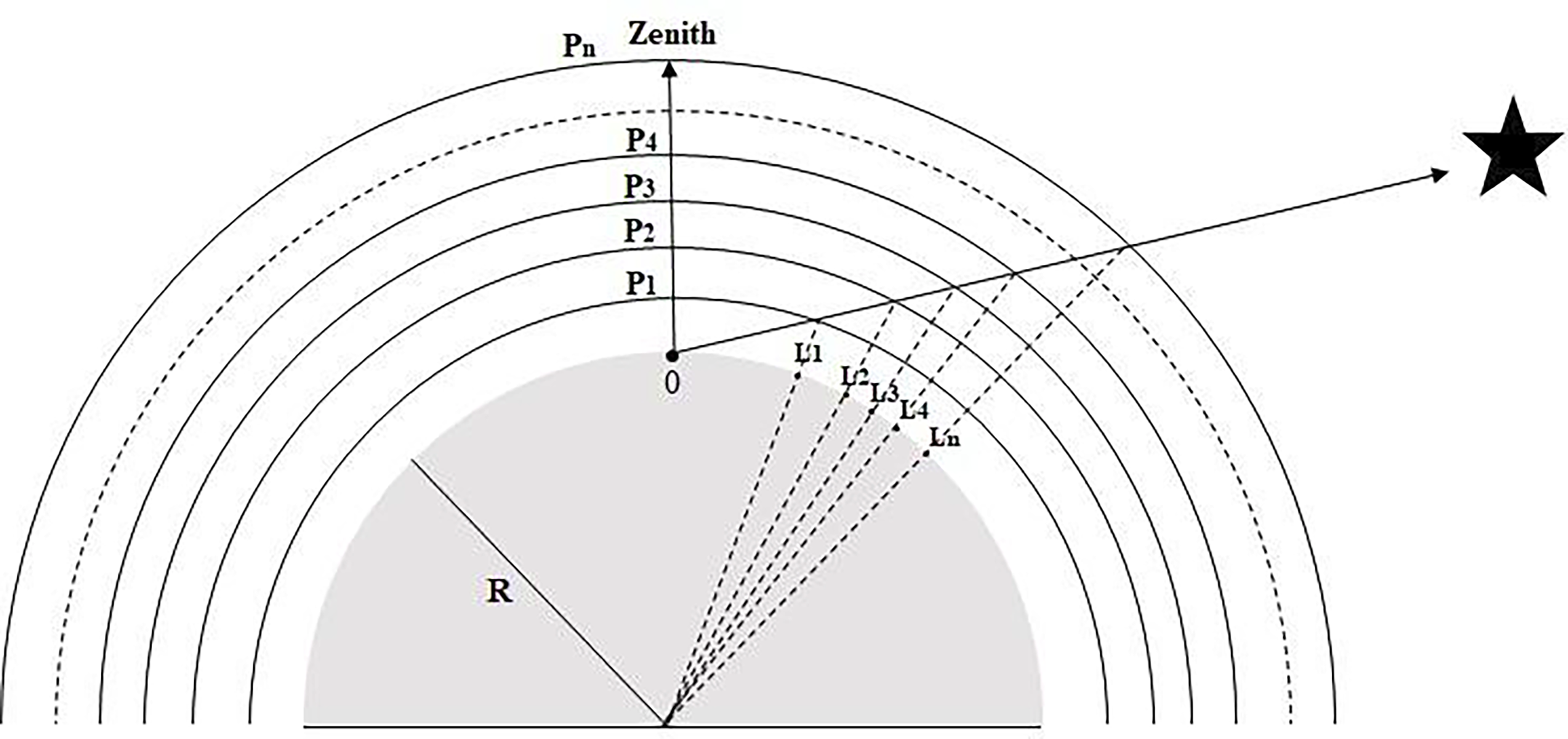}}
    \caption{Illustration on how the projection points $L_{i}$ are located. The PWV is estimated from the observer O along the line up to the pressure level $P_{n}$.}
    \label{fig:Projection}
\end{figure}

In this section the novel method to compute PWV along the line of sight to the target in the sky is described. Computing PWV at the zenith is the easiest task because at all times the values of temperature and relative humidity are found in the same coordinates on the GOES-R data files, but estimating PWV along the line of sight to the celestial target requires additional computation. The idea behind our procedure is to compute an inclined column along the line of sight to the target instead of doing so vertically above the location. Figure \ref{fig:Projection} illustrates this concept (note that the scales are exaggerated for better understanding). Let O be an observation point on Earth looking at an arbitrary target in the sky. Consider each pressure level as a circular shell concentric to Earth. The line of sight to the target intercepts the $i$\textsuperscript{th} pressure layer $P_{i}$ at a certain point. Now, if we connect this interception point with a straight line to the centre of the Earth we obtain a point $L_{i}$ on the surface of the Earth which gives us the projection of the point where the line of sight intercepts the $i$\textsuperscript{th} pressure level.  If a target is fixed at the zenith, then all projection points from all pressure levels are located at the observation's location, but for any other location on the sky, the latitude and longitude coordinates of the projection points have to be computed in order to retrieve temperature and relative humidity data at each relevant pressure level.

To locate the projected places on Earth, the line of sight has to be specified from the point of observation. Horizontal celestial coordinates are the most suited for this, since a target's coordinates are given by the altitude (or also called elevation) angle $\alpha$, azimuth $\delta$ and time in the observation frame of reference \citep[see][chap. 2]{smart}. Assuming that the point of observation lies on a plane tangential to Earth's surface, the projection point at a certain height can be computed with Pythagoras' theorem on euclidean geometry. This assumption of a flat surface is valid for the scales considered here. Estimating PWV up to a pressure level of 300 hPa corresponds to a height of approximately 11 km. Constraining observations to elevation angles above 30 degrees, the maximum projected distance is 19.05 km in euclidean geometry, while the arc length on spherical geometry from the observation point to the projection point equals 18.97 km. The difference is about 81.64 meters, which gives the maximum error by this method. Considering that GOES-R spatial resolution is 10 km, this error is negligible and thus assuming a flat surface is valid.

Once the geodesic distance $\Delta d$ to the observation point is computed, the azimuth angle is used to estimate the distance along longitude $\Delta d_{lon}$ and latitude $\Delta d_{lat}$ lines. Finally the angle subtended from the centre of the Earth between the observation point and the projection point is computed for both latitude and longitude and then added to the coordinates of the observation point which yields the coordinates of the projection point ($lat_{projection}$ and $lon_{projection}$). These coordinates are used to locate the temperature and relative humidity data on the GOES-R data files. All equations are listed below:
\begin{IEEEeqnarray}{rll}
    \Delta d &=& \frac{h}{tan(\alpha)} \label{eq:projections1} \\
    \Delta d_{lat} &=& \Delta d \cdot cos(\delta) \label{eq:projections2} \\
    \Delta d_{lon} &=& \Delta d \cdot sin (\delta) \label{eq:projections3} \\
    lat_{projection} &=& lat_{observer}+arctan\left(\frac{\Delta d_{lat}}{R_{Earth}}\right) \label{eq:projections4} \\
    lon_{projection} &=& lon_{observer}+arctan\left(\frac{\Delta d_{lon}}{R_{Earth}}\right), \label{eq:projections5}
\end{IEEEeqnarray}
where all angles are given in radians, $R_{Earth}$ is the radius of Earth and $h$ the height computed from the pressure via the barometric formula (see Eq. \ref{eq:barometric}) . 

\section{Precipitable water vapour program}
\label{section:fyodor}

We present \texttt{fyodor}, a Python package for computing precipitable water vapour above a location or along the line of sight for arbitrary astronomical sources from publicly available GOES-R data products. \texttt{fyodor} is built on several open source packages including: \texttt{python-netCDF4}, \texttt{NumPy} \citep{harris}, \texttt{matplotlib} \citep{hunter}, \texttt{AstroPy} \citep{astropy:2013,astropy:2018}, \texttt{astroplan} \citep{astroplan2018}, and \texttt{bs4} \citep{richardson2007beautiful}. 
Here we describe the process from downloading the files until the retrieval of PWV. Each moisture and temperature file provided by NOAA has to be locally stored. Users can either search and request data on given dates and times or they can create a subscription for delivery of specific GOES-R products. The user must download each file individually. The program works with GOES-16 and GOES-17 imaging data. For a complete day on the Full Disk region, every 10 minutes a new file with data is available, making a total of 144 files of temperature data and 144 files of moisture data. Each file has a size of approximately 240 MB, corresponding to 70 GB of data for a complete day. However, GOES-R satellites take measurements of the continental United States (CONUS) region every 5 minutes, which doubles the amount of files but reduces the size of each one of them to 30 MB, equivalent to 17 GB for a complete day. If the user prefers to download TPW, each file for the Full Disk and CONUS region has a size of 1.4 MB and 0.28 MB respectively, corresponding to 201.6 MB and 80 MB for a complete day, respectively.  

The name of a typical GOES-R file contains the abbreviation of the product (in our case LVMP, LVTP and TPW), the region of observation (F for Full Disk and C for CONUS), satellite (G16 or G17) and the starting time of measurement in following order: year, number of day, hour, minutes and seconds. Files obtained by a subscription have the number of processing order at the beginning of the name. Here we will show an example with four files, two moisture and two temperature measurements. One of the files was retrieved via subscription for clarity on how the different steps of the program work.

\subsection{\texttt{download\_nc()}: Download the files}
\label{section:downloadnc}

To avoid manually downloading each file, one can call the \texttt{download\_nc} function. It takes three string arguments: the website provided by NOAA with the downloadable files, the directory where a new folder containing the files should be created and the day on which the measurements were taken in format: ``dd mm yyyy'':

\begin{lstlisting}[caption=Download .nc files][H]
>>> import os
>>> from fyodor import download_nc
>>> url = 'https://*.class.noaa.gov/download/*/*'
>>> path = '/User/example' 
>>> day = '01 12 2019'
>>> download_nc(url, path, day)
All files downloaded!
>>> print(os.listdir())
1234.OR_ABI-L2-LVMPF-M6_G16_s20193350120202.nc
OR_ABI-L2-LVMPF-M6_G16_s20193350110202.nc
OR_ABI-L2-LVTPF-M6_G16_s20193350110202.nc
OR_ABI-L2-LVTPF-M6_G16_s20193350120202.nc
\end{lstlisting}

\texttt{download\_nc} was developed specifically for the design of NOAA's Comprehensive Large Array-data Stewardship System (CLASS) website, consisting of a table with every netCDF4 file requested by the user. As long as NOAA's CLASS website remains unchanged the routine will work. The caveat of this method is that due to an occasional slow response by the website or slow internet connection the script tends to time out and fail, forcing the user to run it again. It is worth mentioning that the code does not differentiate between files corresponding to the Full Disk and CONUS region. The user has to make sure to put the different datasets on separate folders.

\subsection{\texttt{rename\_nc()}: Rename subscription files}
\label{section:renamenc}

Since the name of each file contains the time of measurement, they can be sorted in chronological order. However, data obtained from a subscription have at the beginning the number of processing order. If the file name is not renamed, sorting the files will not necessarily order them chronologically. To delete the first string of numbers of all files in a specific folder one can call \texttt{rename\_nc}, giving the directory as the argument:

\begin{lstlisting}[caption=Rename .nc files][H]
>>> import os
>>> from fyodor import rename_nc
>>> path = '/User/example/Data PWV 01 19 2019' 
>>> rename_nc(path)
>>> print(os.listdir())
OR_ABI-L2-LVMPF-M6_G16_s20193350120202.nc
OR_ABI-L2-LVMPF-M6_G16_s20193350110202.nc
OR_ABI-L2-LVTPF-M6_G16_s20193350110202.nc
OR_ABI-L2-LVTPF-M6_G16_s20193350120202.nc
\end{lstlisting}
Note that the name of the first file was changed by removing the string "1234".

\subsection{\texttt{pwv()}: Compute PWV at the zenith}
\label{section:pwv()}

The package \texttt{netCDF4} reads the files and lists all variables, range and dimensions contained in the file. All necessary Earth parameters (such as semi-major axis) are assigned here into variables reading out just the first file in the folder.

Each file contains 101 standard pressure levels starting at $1100$ hPa up to $0.005$ hPa and these are the same for all GOES-16 products containing the pressure variable. For this reason, the script assigns the pressure variable just once out of the first file. 

The data files contain time variables represented in seconds since J2000 epoch (2000-01-01 12:00:00 UTC). The script converts to human readable time in UTC format. 

In order to compute the height as a function of pressure, hydrostatic balance in the atmosphere is assumed. The height depends on the pressure ($P$), the universal gas constant ($R$), acceleration due to Earth's gravity ($g$), standard atmospheric pressure ($P_{0}$) and a typical surface temperature ($T_{0}$) of 288 K \citep{wells}. We also assume a constant lapse rate ($L$). The height has to be considered as an approximation but it is necessary to compute the PWV along the line of sight to a target in the sky:
\begin{IEEEeqnarray}{rll}
\label{eq:barometric}
h = \frac{T_{0}}{L}\left(\left(\frac{P}{P_{0}}\right)^{-L \cdot R/g}-1\right)
\end{IEEEeqnarray}
Although the variation on gravitational acceleration and the gas constant are negligible from the surface to 300 hPa, the temperature changes considerably over space and time. The limitation of the formulation used in the code is that the reference temperature of 288 K is the same for any location for any time of the year.  

The \texttt{pwv} function requires several arguments. As for all other objects, the directory containing the files is necessary. The observing location is required and accepts three different inputs: "Cerro Paranal", "San Pedro Martir" and "Other". The first two will set automatically Cerro Paranal or San Pedro M\'artir's observatory latitude and longitude. Alternatively, "Other" gives the user the option to enter a valid value for latitude (between -81.3281 and  81.3281 degrees) and longitude (between -156.2995 and 6.2995 degrees), defined in the GOES-R documentation \citep{PUG1}. All relevant coordinates located on Earth are transformed into scan angles according to Eq. (\ref{eqn:scanangles}) with its auxiliary equations (\ref{eqn:scanangles2}) and (\ref{eqn:scanangles3}).

Next, the function needs the pressure level boundaries. As already discussed in the Introduction, above 300 hPa the contribution to PWV is very small, being reasonable to set it as a cutoff. The lower boundary (in height) can be set to the standard atmospheric pressure or ideally the surface pressure at the location of interest. In Cerro Paranal the mean atmospheric pressure on the surface is 750 hPa \footnote{\url{https://www.eso.org/sci/facilities/paranal/astroclimate/site.html}} while in San Pedro M\'artir it is 727 hPa \footnote{\url{http://tango.astrosen.unam.mx/clima/almanac_Plus.htm}}.

To compute PWV above the location, one has to set the argument \texttt{line\_of\_sight}='zenith'. Finally, \texttt{plot}=True displays a plot with the results and saves a .png figure in the directory, while \texttt{csv}=True will generate and save a .csv file with the time of measurement and PWV value. Below is a complete example at Cerro Paranal:

\begin{lstlisting}[caption=Compute PWV above Cerro Paranal][H]
>>> from fyodor import pwv
>>> directory = '/Users/example/Data PWV 01 19 2019' 
>>> location = 'Cerro Paranal'
>>> P_min = 750
>>> P_max = 300
>>> line_of_sight = 'zenith'
>>> date, water = pwv(directory, location, P_min, P_max, line_of_sight, RA=None, Dec=None, plot=False, csv=False)
>>> date
['2019-12-01 01:15:05', '2019-12-01 01:25:05']
>>> water
array([2.92674977, 2.89279106])
\end{lstlisting}

\subsection{\texttt{pwv()}: Compute PWV along the line of sight}
\label{section:pwv()2}

Alternatively, to obtain PWV along the line of sight to a target, one has to set \texttt{line\_of\_sight}='target'. The coordinates of the target in equatorial coordinates have to be given. In Sect. \ref{section:Precipitable water vapour along the line of sight to the target} the equations were derived making use of horizontal celestial equations, but instead the user is still asked to input Right ascension (RA) and Declination (Dec) values. The reason relies on the fact that equatorial coordinates require less input information, since RA and Dec are constant numbers (during a long period of time) and computationally it is much easier to transform from RA and Dec to Altitude and Azimuth than typing these angles and time of measurement and then compute the angles for each time of measurement of the satellite (i.e. each data file). The \texttt{astropy} package carries out the transformation and assigns the altitude and azimuth angles to each time of measurement. At the moment the code supports only RA and Dec given in degrees. While it is customary to enter RA in hour angles, \texttt{astropy} is a suited program to make conversions and the option to input different units will be implemented soon.

Another advantage of having the coordinates of a target given by its altitude and azimuth is that one can immediately check if the target is above the horizon by considering altitude angles above zero degrees. In this study an additional constraint was set, because usual observations are done down to a certain elevation: we only consider elevation angles above 30 degrees. The code stores in a variable the indices where this condition is fulfilled. 

The latitude and longitude coordinates of the projection points for each file and each pressure level (height) are computed as described by Eqs. (\ref{eq:projections1}) to (\ref{eq:projections5}). They are transformed into scan angles and its horizontal and vertical index number of the pixel in the files are retrieved by searching for the closest available gridpoint in the dataset. For each of these indices and each pressure level the corresponding temperature and relative humidity values are retrieved.

To compute the amount of PWV given by Eq. (\ref{eqn:PWV3}) a discrete numerical integration based on the trapezoidal method is used. For each file (i.e. for each time of measurement) the integral is computed between the pressure level boundaries.

The example below displays a target with coordinates RA=0, Dec=0 observed from San Pedro M\'artir:

\begin{lstlisting}[caption=Compute PWV along the observing line of sight as observed from San Pedro M\'artir][H]
>>> from fyodor import pwv
>>> directory = '/Users/example/Data PWV 01 19 2019' 
>>> location = 'San Pedro Martir'
>>> P_min = 727
>>> P_max = 300
>>> line_of_sight = 'target'
>>> RA = 0
>>> Dec= 0
>>> date, water = pwv(directory, location, P_min, P_max, line_of_sight, RA, Dec, plot=False, csv=False)
>>> date
['2019-12-01 01:15:05', '2019-12-01 01:25:05']
>>> water
array([2.57504235, 2.61801297])
\end{lstlisting}

\subsection{\texttt{tpw()}: Compute PWV using the TPW product}
\label{section:tpw()}

If the user prefers to use the Total Precipitable Water product provided by NOAA, the program can compute PWV using the \texttt{tpw} function, which is a simpler version than \texttt{pwv}. It needs the working directory and the locations as described for \texttt{pwv}. We show the working of this function using the corresponding TPW files for the same time of measurements as the examples above at Cerro Paranal:

\begin{lstlisting}[caption=Compute PWV using the TPW product above Cerro Paranal][H]
>>> from fyodor import tpw
>>> directory = '/Users/example/Data PWV 01 19 2019' 
>>> location = 'Cerro Paranal'
>>> date, water = tpw(directory, location, plot=False, csv=False)
>>> date
['2019-12-01 01:15:05', '2019-12-01 01:25:05']
>>> water
[6.841148, 6.8823504]
\end{lstlisting}

\section{Results and discussion}
\label{section:results}

We analysed the amount of precipitable water vapour estimated by the program for two weeks of consecutive data from 1 December to 15 December 2019 as well as 1 March to 15 March 2020. The first section contains the results of PWV above Cerro Paranal computed with the function \texttt{pwv} compared with data provided by LHATPRO and the TPW product. In Sect. \ref{section:PWVzenithSPM} we present the results of PWV above the observatory located in San Pedro M\'artir. We computed PWV between 750 hPa and 300 hPa pressure levels in Cerro Paranal and between 727 hPa and 300 hPa in San Pedro M\'artir. One of the reasons we chose this upper limit is because the PWV in the TPW product is computed until that level and thus provides better conditions for comparison. The measurements were taken every 10 minutes by GOES-16. All times are given in UTC. The results of PWV at zenith contain the complete number of available data excluding missing data, but the results of PWV along the line of sight to a target in the sky in Sect. \ref{section:PWVls} consists of observations constrained to elevation angles above 30 degrees. Finally, in Sect. \ref{section:PWV above 300 hPa} we briefly discuss the contribution to PWV above the pressure level of 300 hPa. From now on, when we refer to \texttt{fyodor}, we mean the function \texttt{pwv} which is its main feature, while the results using the TPW product were retrieved using \texttt{fyodor}'s function \texttt{tpw}. To clarify, when we mention \texttt{fyodor}'s measurements or results, we mean the PWV from \texttt{fyodor} code using GOES data.

\subsection{Comparison of PWV above Cerro Paranal with LHATPRO data and TPW}
\label{section:ComparisonfyodorLHATPROtpw}

\begin{figure*}%
    \centering
    \subfloat{{\includegraphics[width=8cm]{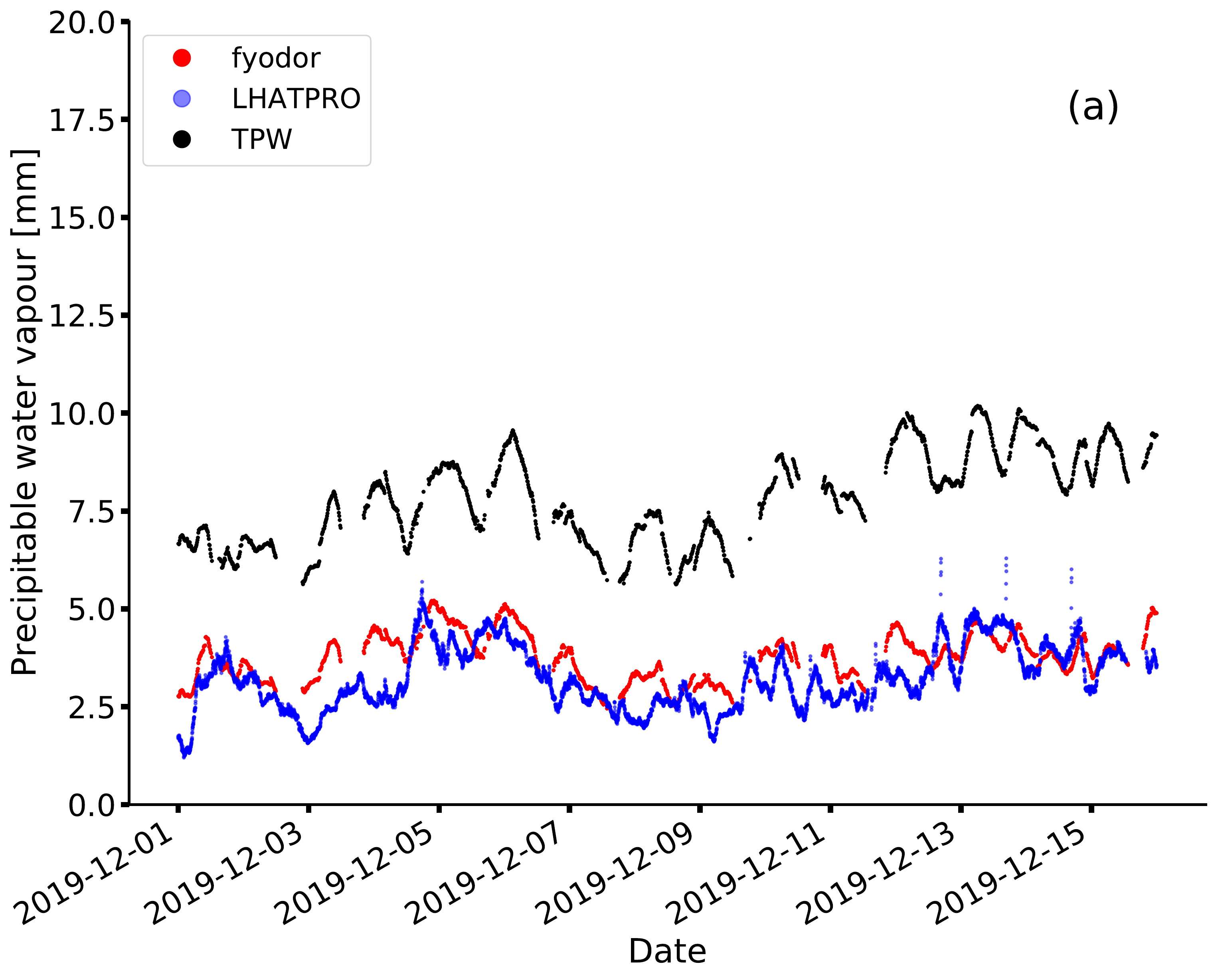} }\label{fig:CPzenithdecember}}%
    \subfloat{{\includegraphics[width=8cm]{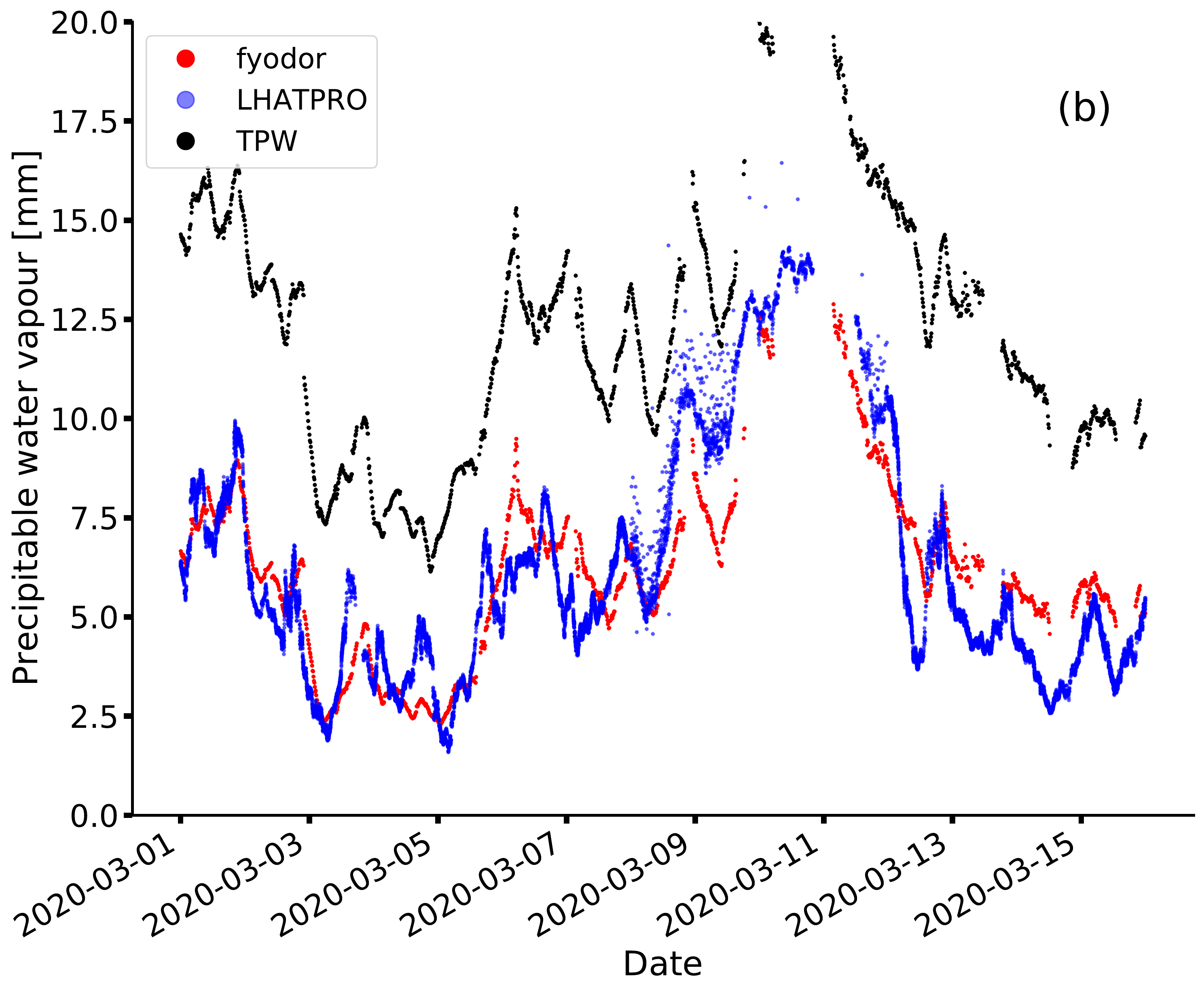} }\label{fig:CPzenithmarch}}%
    \caption{PWV above Cerro Paranal. Red points are \texttt{fyodor}'s measurements taken between 750 hPa and 300 hPa every 10 minutes using GOES-16 imagery data of the Full Disk region. Blue correspond to LHATPRO and black to the TPW product using GOES-16 imagery data from (a) 2019-12-01 to 2019-12-15 and (b) 2020-03-01 to 2020-03-15.}%
    \label{fig:CPzenith}%
\end{figure*}

To asses the code's performance, we compared PWV data with values measured by LHATPRO, a ground-based radiometer installed at Cerro Paranal. LHATPRO measurements are publicly available to download\footnote{\url{http://archive.eso.org/wdb/wdb/asm/lhatpro_paranal/form}}. We also compared the results of \texttt{fyodor} with the TPW product. In Fig. \ref{fig:CPzenith} each panel presents two weeks of measurements. Red dots correspond to PWV from GOES data computed by \texttt{fyodor}, blue to LHATPRO and black to TPW. In December (Fig. \ref{fig:CPzenithdecember}), the difference between the highest and lowest value, called peak to peak value, of 2.8 mm given by \texttt{fyodor} is smaller than LHATPRO (5 mm) and TPW (4.6 mm), indicating that the method may be less sensitive to changes than other instruments. During the first two weeks in March (Fig. \ref{fig:CPzenithmarch}) the amount of PWV is considerably higher than in December, showing a seasonal variability. The change in PWV during a single day in March is also highly variable, ranging between 1.1 mm and 4.4 mm difference within a day. 

For most of the measurements in December the PWV given by \texttt{fyodor} is higher than those by LHATPRO, agreeing between each other within 3$\sigma$. In March, both dataset are similar at a significance of 0.25$\sigma$. By eye it is clear that the trend is similar. Most of the variable features are well captured by both \texttt{fyodor} and LHATPRO. However, some drops in PWV measured by LHATPRO, especially on 4 and 8 December 2019, are inverted in the estimation given by \texttt{fyodor}, resulting in an increase of PWV. For higher values both datasets seem to agree better with each other. 

Analysing all our data it seems that \texttt{fyodor}'s differences in values over time are less subtle than LHATPRO's. While the general behaviour is similar between the two, LHATPRO appears to be more sensitive to PWV variations. The reason behind the different values relies on the fact that LHATPRO takes measurement in a range between 183 and 191 GHz \citep{querel} -- which corresponds to strong absorption lines \citep{matsushita} --, 51 to 58 GHz and in the range of 10 $\mu$m, while GOES-16 data is taken at different bands centred between 3.89 and 13.27 $\mu$m. The absorption coefficient of water vapour was checked in the HITRAN database\citep{Gordon} for these different regions. It is highly variable and not identical for the range where LHATPRO and GOES-16 obtain measurements \citep[see][]{Tennyson, Taylor}, which explains the absolute value difference in PWV.  
Another reason for the difference relies on the fact that LHATPRO is an \textit{in situ} instrument, guaranteeing that the measurements correspond to Cerro Paranal, while GOES-16 has a spatial resolution of 10 km, meaning that every measurement represents an average value over an area of 10 km x 10 km horizontal size. Thus, temperature and relative humidity values do not necessarily correspond to the actual values on Cerro Paranal, but to those at the nearest gridpoint on the dataset.

Comparing our results with the TPW product, both datasets also follow the same trend, capturing every drop or increase in PWV. TPW's values seem shifted up by an offset and stretched by a factor. The reason why both datasets seem so similar excluding the shift relies on the fact that the TPW end product is generated by the same algorithm that produces the Legacy Vertical Temperature (Moisture) Profile and is derived from the moisture profile. However, TPW gives an integrated value of PWV between the surface and 300 hPa. The offset seen in the TPW product is a result of taking into account pressure levels that may not correspond to the surface pressure on a certain location. This is especially true for high altitude locations. Running some tests with \texttt{fyodor}, we found that to match the offset of the TPW product the surface pressure at Cerro Paranal should be approximately 950 hPa, while the actual mean surface pressure is approximately 750 hPa. This leads to an overestimated value of PWV by the TPW product. 

\subsection{PWV above San Pedro M\'artir}
\label{section:PWVzenithSPM}

\begin{figure*}%
    \centering
    \subfloat{{\includegraphics[width=8cm]{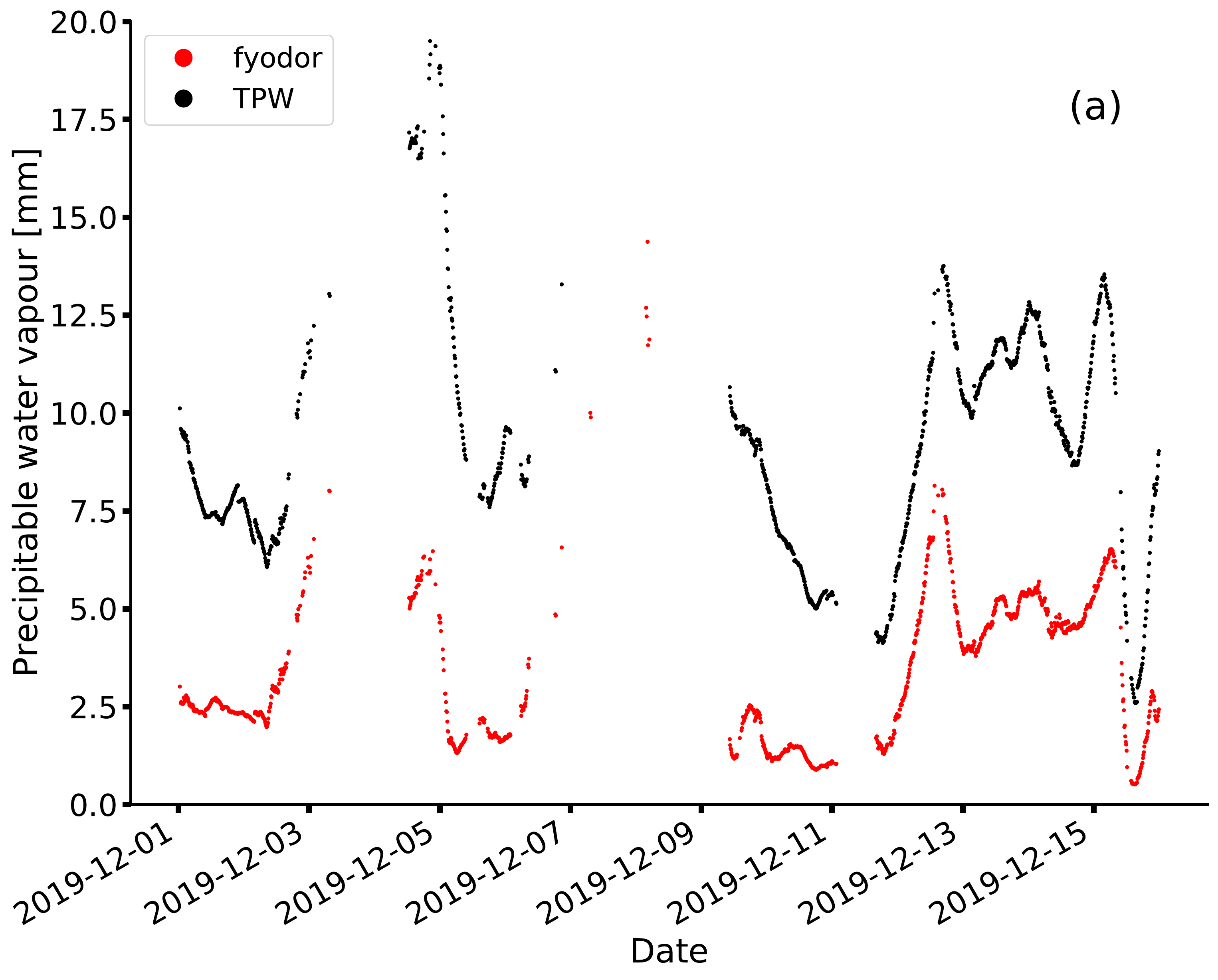} }\label{fig:SPMzenithdecember}}%
    \subfloat{{\includegraphics[width=8cm]{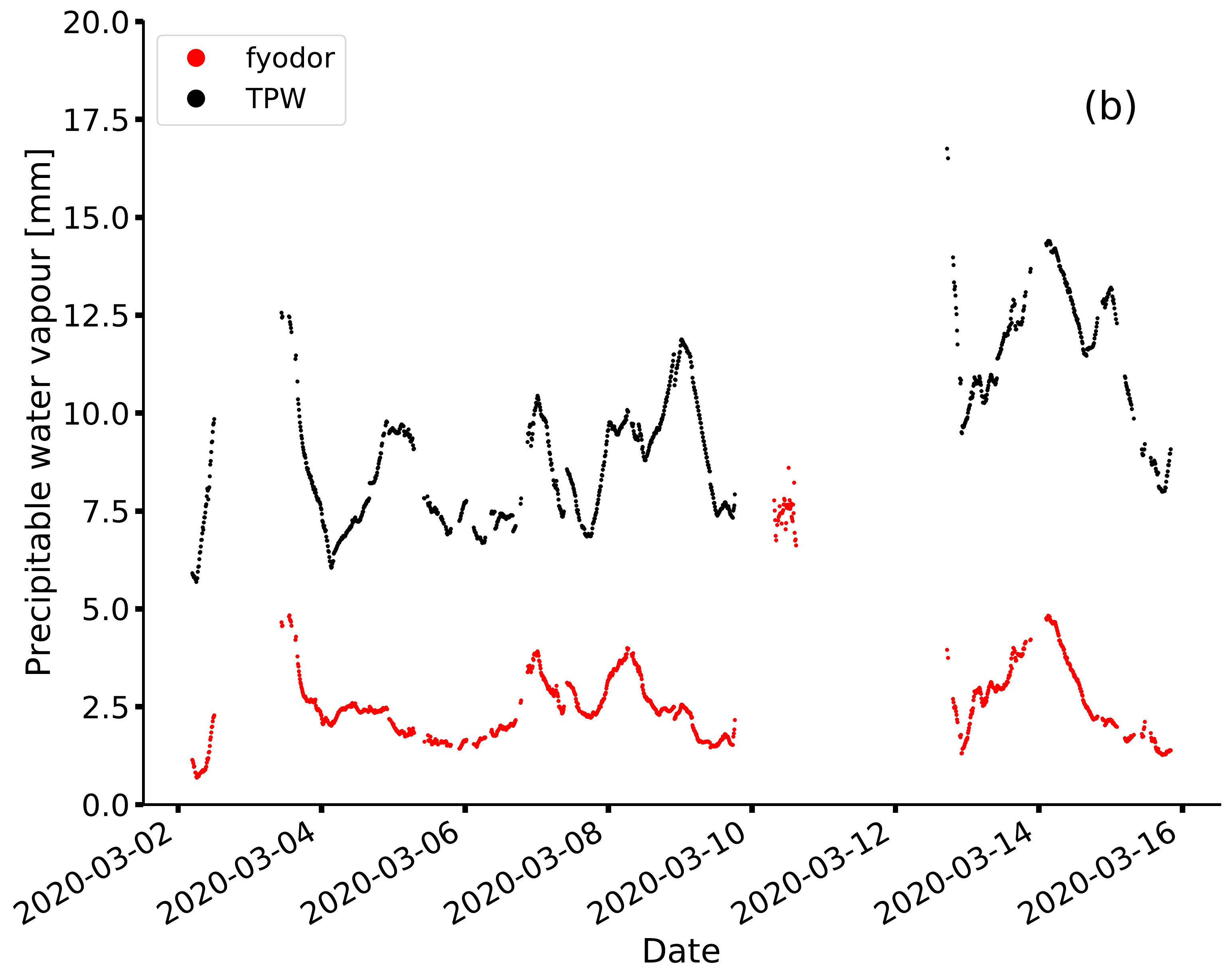} }\label{fig:SPMzenithmarch}}%
    \caption{PWV above San Pedro M\'artir. Red points are \texttt{fyodor}'s measurements taken between 727 hPa and 300 hPa every 10 minutes using GOES-16 imagery data of the Full Disk region and black to the TPW product using GOES-16 imagery data from (a) 2019-12-01 to 2019-12-15 and (b) 2020-03-01 to 2020-03-15.}%
    \label{fig:SPMzenith}%
\end{figure*}

We computed the amount of PWV above San Pedro M\'artir for the same dates using \texttt{fyodor} and TPW. In Figs. \ref{fig:SPMzenithdecember} and \ref{fig:SPMzenithmarch} there are some notable gaps of missing data between 3 and 4 December 2019 and between 10 and 12 March 2020, probably due to partially cloudy or overcast conditions, since the values of PWV before and after the gaps are relatively high. There is no evident seasonal variability between these months.

The difference between both measurements is analogous to the analysis above Cerro Paranal in Sect. \ref{section:ComparisonfyodorLHATPROtpw}. The trend is similar, but TPW overestimates the amount of PWV compared to \texttt{fyodor}. However, there are some PWV variations in the TPW product that our program does not recognise or the difference between values is very narrow. The most drastic event occurred on 5 March 2020, where a considerable increase in PWV recorded by TPW is just a small bump in \texttt{fyodor}'s results. We can see similar results on 9 and 15 March 2020, where there is even no evidence of the increase in PWV. Since TPW is a quality-checked end product, it could be that it undergoes extra processing and computation than the products used by \texttt{fyodor}. 

When computing PWV above San Pedro M\'artir using GOES-16 imagery data the user can make use of one of the regions covering the location: Full Disk and CONUS. The Full Disk dataset has a temporal resolution of 10 minutes, while new data is available in the CONUS dataset, every 5 minutes. However, during a test we found discrepancies between the values in the datasets. The measurements are independent from each other and usually they do not overlap in time of measurement. However, some of the CONUS values may be extracted from the Full Disk measurement, while others are independent scans \citep{PUG5}. For full coverage of the American region and consistency in the measurements, we recommend using the Full Disk region dataset.

\subsection{PWV along the observing line of sight from Cerro Paranal and San Pedro M\'artir}
\label{section:PWVls}

\begin{figure*}%
    \centering
    \subfloat{{\includegraphics[width=8cm]{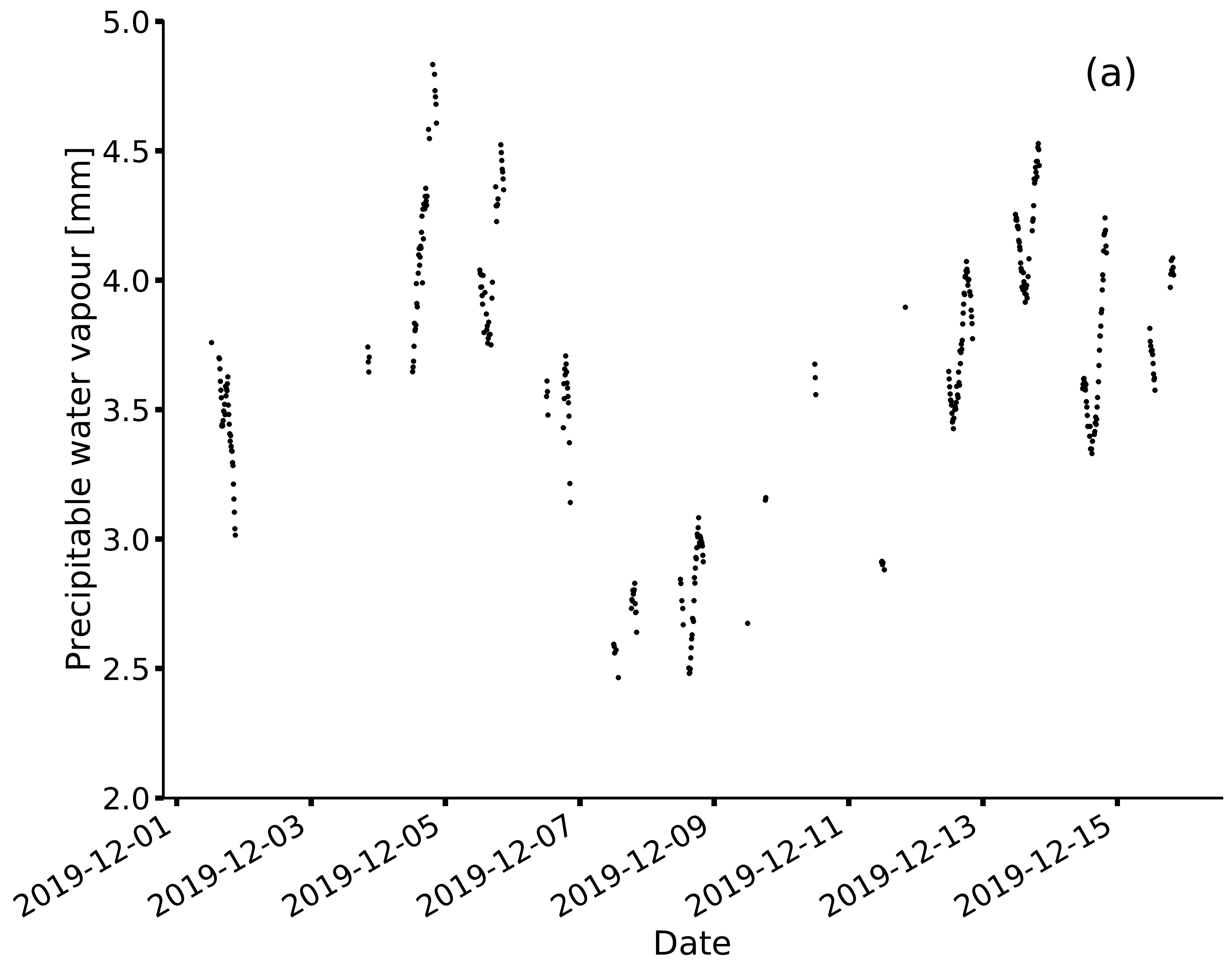} }\label{fig:CPlosdecember}}%
    \subfloat{{\includegraphics[width=8cm]{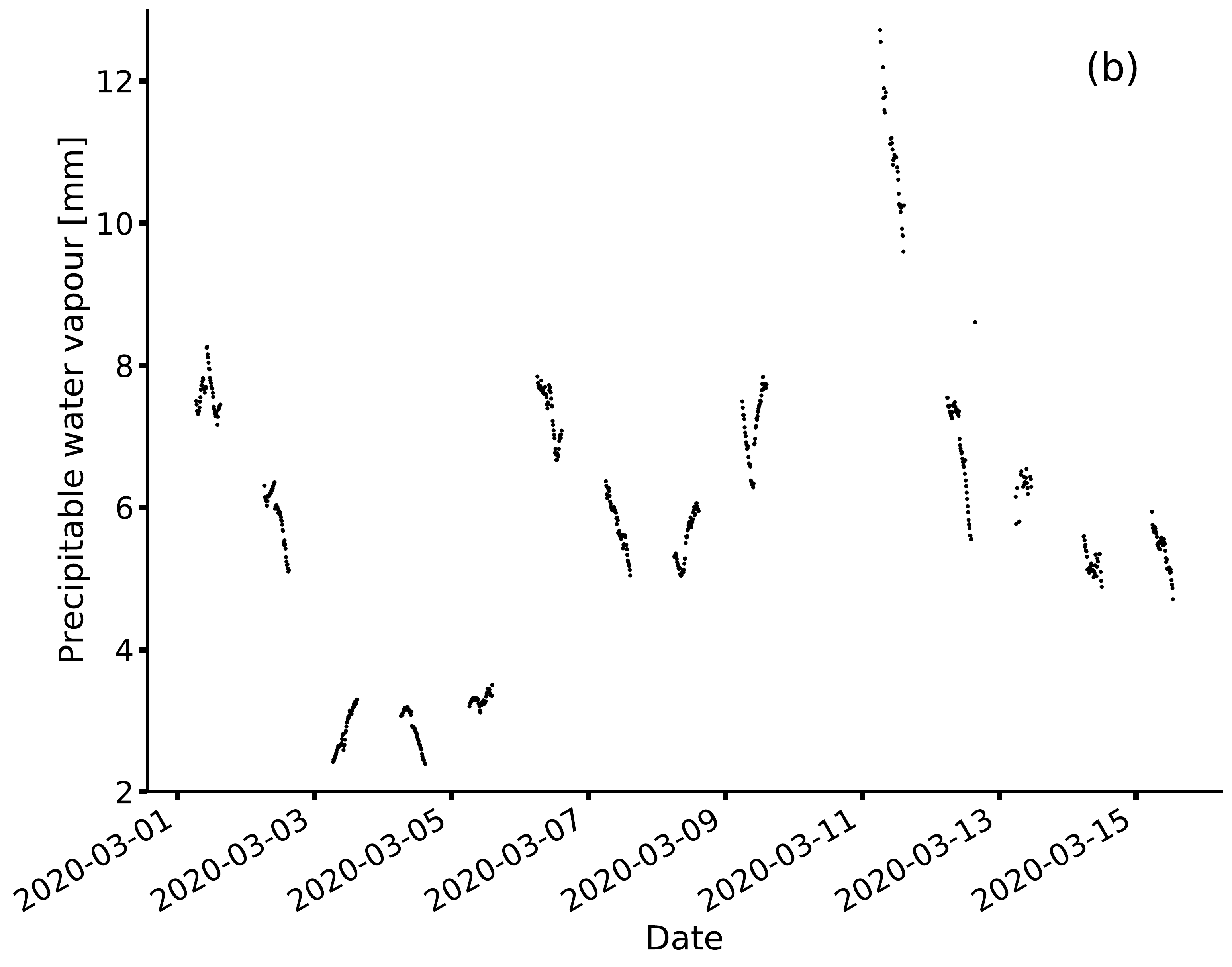} }\label{fig:CPlosmarch}}%
    \caption{PWV along the line of sight to a test target with coordinates $RA = 250$ degrees and $Dec = -20$ degrees observing from Cerro Paranal between 750 hPa and 300 hPa every 10 minutes using GOES-16 imagery data of the Full Disk region from (a) 2019-12-01 to 2019-12-15 and (b) 2020-03-01 to 2020-03-15. Data available only when the target was visible above an elevation angle of 30 degrees. Note the difference in the vertical axes limit.}%
    \label{fig:CPlos}%
\end{figure*}

To prove the estimation of PWV along the line of sight, we entered in the program a toy test target with coordinates Right Ascension (RA) $= 250$ and Declination (Dec) $= -20$ as seen from the observatory in Cerro Paranal. By definition, right ascension can be related to the longitude on the great circle, while declination corresponds to latitude. In general, in order to compute PWV along the line of sight, the code is ingested with two different sets of coordinates: The observing location and the coordinates of the target in the sky. For this particular example the test target has similar coordinates than the observatory. In this fashion, once in a 24h cycle it passes very close to the observatory's zenith. The results (Figs. \ref{fig:CPlosdecember} and \ref{fig:CPlosmarch}) show PWV values for time intervals when the target was visible (above an elevation angle of 30 degrees). The PWV change during an observing window in December ranges between 0.2 mm on 6 December 2019 to 1.2 mm on 4 December 2019, while in March it ranges between 0.4 mm on 5 March 2020 to 3.1 mm on 11 March 2020. The trend is similar than above the location (see Figs. \ref{fig:CPzenithdecember} and \ref{fig:CPzenithmarch}) as well as the range of values, but there are clearly fewer data points. The reason is that by constraining to observations above 30 degrees altitude the maximum projection distance is around 20 km, as discussed in Sect. \ref{section:Precipitable water vapour along the line of sight to the target} but the dataset has a resolution of 10 km. This means that the maximum change of pixels containing data points in the ABI Fixed grid equals two pixels. In other words, temperature and moisture data of projection points is retrieved quite close to the observation point if not exactly at it.

\begin{figure*}%
    \centering
    \subfloat{{\includegraphics[width=8cm]{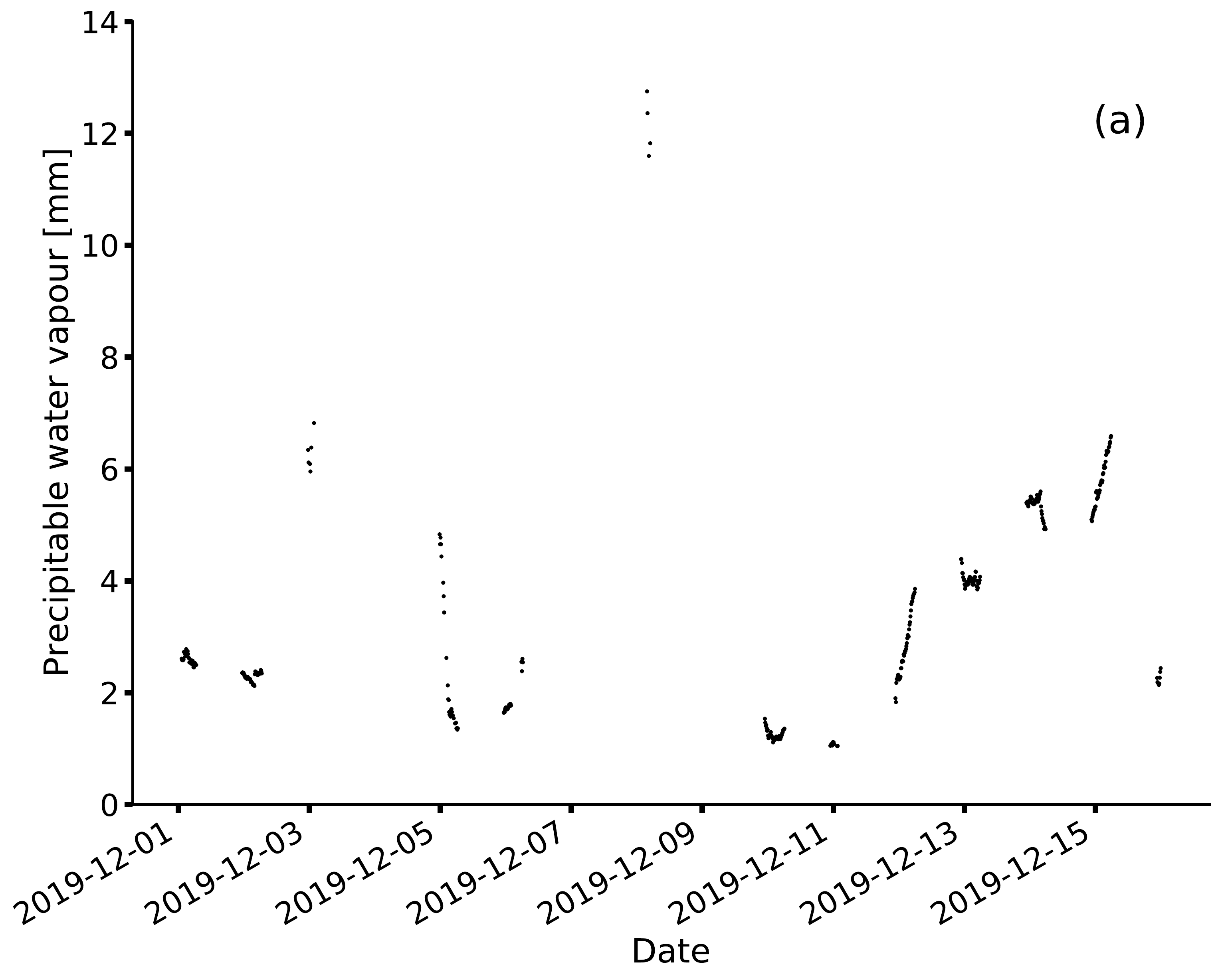} }\label{fig:SPMlosdecember}}%
    \subfloat{{\includegraphics[width=8
    cm]{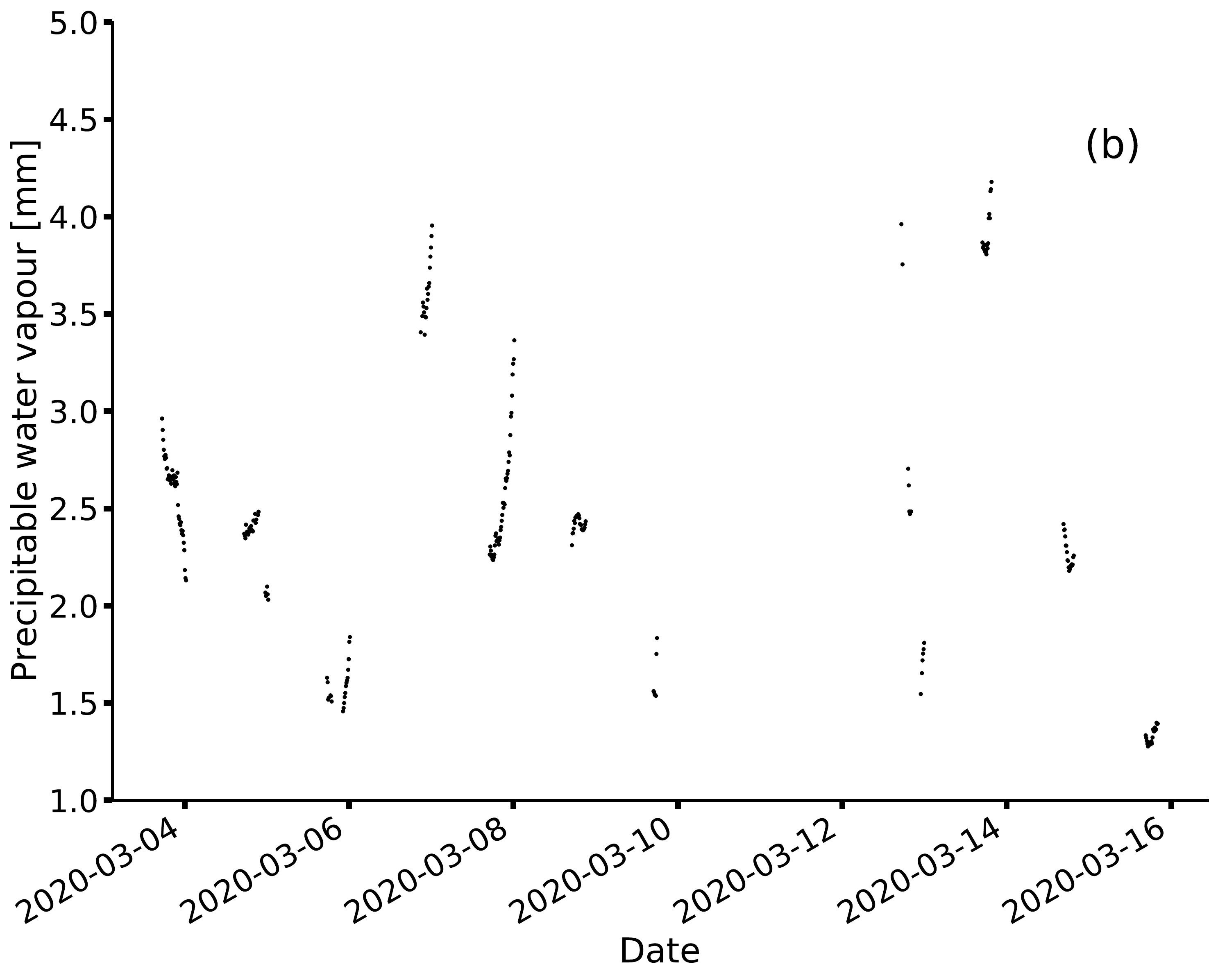} }\label{fig:SPMlosmarch}}%
    \caption{PWV along the line of sight to a test target with coordinates $RA = 0$ degrees and $Dec = 0$ degrees observing from San Pedro M\'artir between 727 hPa and 300 hPa every 10 minutes using GOES-16 imagery data of the Full Disk region from (a) 2019-12-01 to 2019-12-15 and (b) 2020-03-01 to 2020-03-15. Data available only when the target was visible above an elevation angle of 30 degrees. Note the difference in the vertical axes limit.}%
    \label{fig:SPMlos}%
\end{figure*}

Next we computed the PWV from San Pedro M\'artir along the line to a target with coordinates $RA = 0$ degrees, $Dec = 0$ degrees, shown in Figs. \ref{fig:SPMlosdecember} and \ref{fig:SPMlosmarch}. During a single observing window of approximately 8 hours, the PWV varies as small as 0.2 mm on 2 December 2019 to 3.4 mm on 5 December 2019. In March, the smallest change during an observation is 0.3 mm on 9 March 2020, while the biggest change of 1.2 mm occurs on 8 March 2020. This results reinforce the purpose for which \texttt{fyodor} was developed, that is, act as a supporting tool monitoring PWV during astronomical observations. 

Given that the instrumentation uncertainty in Relative humidity is 18\% and 1 K in Temperature, our first estimates give a percentage uncertainty of around 27\%. One has also to take into account the error we induce by the numerical integration, the height approximation via barometric formula and the formulation of the saturation vapour pressure.

\subsection{Contribution to PWV above 300 hPa}
\label{section:PWV above 300 hPa}

In Sect. \ref{section:introduction} we mentioned that at heights above a pressure level of 300 hPa the contribution to PWV is very small, as also stated by \citet{erasmus3}. To make sure, we analysed our data above Cerro Paranal and San Pedro M\'artir between 300 hPa and 0,005 hPa, which is the smallest pressure level available in the GOES-16 dataset.

Above Cerro Paranal the amount of PWV is still variable in time, ranging between 0.05 and 0.15 mm. Compared to the precipitable water vapour computed in Cerro Paranal between the surface and 300 hPa, the content above 300 hPa represents between 0,8\% and 1,5\% of it. 

The results above San Pedro M\'artir are similar. All values are found between 0.02 and 0.047 mm, representing 0,87\% and 1,67\% of the amount found between the surface and 300 hPa. Thus, we conclude that the contribution to PWV above 300 hPa is indeed small.

\section{Conclusions}
\label{section:conclusions}

The high variability of precipitable water vapour in the atmosphere presents itself as a serious challenge in the search for small exoplanets orbiting ultra-cool stars. Using GOES-16 satellite imagery data, the Python program developed here has proven to be a promising assisting tool for astronomical observation giving an estimated value of precipitable water vapour above the site. Furthermore, we conclude that our program is correctly tracking the motion of the target, benchmarking the novel method proposed here to compute PWV along the line of sight to the target in the sky during the time of observation.

Computing PWV using satellite-based imagery data has the advantage of being a remote sensing method reasonably inexpensive for the user allowing computations for many locations across the complete American continent. It can be very helpful at observatories within the GOES coverage that do not benefit from  an on-site PWV monitoring tool. Additionally, the temporal resolution of 10 minutes or 5 minutes (depending on dataset) allows to clearly recognise temporal variability during typical observation periods in near real-time. For a single measurement, a gross estimate of the time a user would need when using \texttt{fyodor} to complete the process from downloading GOES data to the time when the amount of PWV is computed and saved in a .csv file is approximately 5 minutes for the Full Disk with a slow internet connection and approximately 30 seconds with a fast connection. In comparison, the whole process using the TPW product takes approximately 5 seconds. However, the time between requesting data to NOAA and receive the link with the files is highly variable. The uncertainty in PWV computed in Cerro Paranal and San Pedro Mártir lies around 27\%. Comparing with data in Cerro Paranal provided by LHATPRO, although the values differ, both follow a similar trend, which is favourable for the reliability on the method developed here. In general the variability on PWV is reasonably alike for both methods and the magnitudes are similar. It is recommended to compare the amount of PWV over a longer period of time to count with a statistical reliable analysis.

The TPW product provided by NOAA has the advantage of being an end product that gives the PWV value. The files are considerably smaller and the running time of the program is also reduced. Users interested in a fast method to retrieve the amount of PWV can use this product along with the function in our program. The caveat is that it only works at the zenith on the location and the pressure level limits are fixed between the surface and 300 hPa, which leads to an overestimated PWV value at high altitude locations. If the user needs PWV between certain pressure levels or to follow a target during observation, he should use the \texttt{pwv} function of \texttt{fyodor} along the temperature and moisture files from GOES-16. As a supporting tool in astronomy, it is not necessary to download a complete day worth of data, the files corresponding to the observation time suffice.

The program was developed with the purpose of monitoring and mitigating the effect of variable PWV on astronomical observations while searching for exoplanets orbiting cool stars, but the program allows other applications to address questions on climate science or forecasting studies. The steps involved to compute PWV lead to the definition of many variables storing useful meteorologic and geographic information, which can prove helpful for modelling. From now on, \texttt{fyodor} program can be used as a supporting tool during observations.

The aim of this project was to develop an open-source program that reliably computes PWV at the request of the user. The software is intended to undergo continuous maintenance and improvements. In the near future, the option to input the altitude and azimuth of a target will be implemented. Also, information regarding quality flags would be helpful to catalogue masked entries. NOAA provides a product containing clear sky mask and cloud mask information, which would surely improve \texttt{fyodor}. However, the number of files required for PWV measurements would increase. Another improvement planned is to automate the process of downloading the dataset. Right now, the user has to request the files of interest on his own at NOAA's website and then run the program to open and process them. Finally, the next big step of improvement is to implement a method that directly corrects noise produced by the presence of precipitable water vapour on observations. At the moment the user can monitor the amount of precipitable water vapour present during his observation at a given location, but the use given to the results is up to the user, that is, correlation or correction have to be done manually. For further research, the program will be put to test with real observations.

\begin{acknowledgements}
We would like to thank the U.S. National Oceanic and Atmospheric Administration (NOAA) for the Legacy Vertical Temperature Profile, Legacy Vertical Moisture Profile and Total Precipitable Water product data. We thank the anonymous referee for the constructive suggestions and comments that improved the manuscript. We are grateful to N. Schanche for her contribution to the code. B.-O. D. acknowledges support from the Swiss National Science Foundation (PP00P2-190080). This work has received support from the Centre for Space and Habitability (CSH) and the National Centre for Competence in Research PlanetS, supported by the Swiss National Science Foundation (SNSF). 
\end{acknowledgements}


\begin{appendix}

\section{Derivation of Eq. \ref{eqn:PWV3}}
\label{section:derivation}
The specific humidity and mixing ratio $w$ of moist air, which is the ratio of the mass of water vapour to the mass of dry air \citep{camuffo} can be related as follow:
\begin{IEEEeqnarray}{rll}
\label{eqn:mrsh}
    q = \frac{w}{1+w}
\end{IEEEeqnarray}
Using the ideal gas equations for water vapour and dry air, considering $e$ as the partial vapour pressure of water and $p_{a}=p-e$ the partial pressure of dry air and p the atmospheric pressure, we can express the mixing ratio as
\begin{IEEEeqnarray*}{rll}
\label{eqn:mrp}
    w &=& \frac{R_{a}}{R_{v}}\frac{e}{p-e} = \frac{M_{v}}{M_{a}}\frac{e}{p-e} =  0.622\frac{e}{p-e} \IEEEyesnumber
\end{IEEEeqnarray*}
where $R_{a}$ and $R_{v}$ are the gas constants for dry air and water, respectively, while $M_{a}=28.966$ is the molecular mass of dry air and $M_{v}=18.016$ is the molecular mass of water.
If we substitute Eq. (\ref{eqn:mrp}) into Eq. (\ref{eqn:mrsh}) we get an expression for the specific humidity as a function of the pressure of water vapour and atmospheric pressure. 
\begin{IEEEeqnarray}{rll}
\label{eqn:shp}
    q = \frac{0.622e}{p-0.378e}
\end{IEEEeqnarray}
Relative humidity $u$ is defined as the ratio between the partial pressure of the vapour and its saturation vapour pressure at the same temperature and pressure. It ranges between 0 and 1 or can also be expressed in percent. We will keep it as a fraction. 
\begin{IEEEeqnarray}{rll}
\label{eqn:rh}
    u = \frac{e}{e_{sat}(T)}
\end{IEEEeqnarray}
Equation (\ref{eqn:rh}) is an useful way to compute the partial pressure of vapour given the relative humidity and saturation vapour pressure, which is a temperature dependent pressure value on which a dynamic equilibrium between evaporation and condensation prevails. The functional form of the saturation pressure is given by the empirical Magnus and Tetens formula \citep{camuffo}:
\begin{IEEEeqnarray}{rll}
\label{eqn:MT}
    e_{sat}(T) = e_{sat}(0) \cdot 10^{aT/\left(b+T\right)}
\end{IEEEeqnarray}
where $e_{sat}(0)=611$ hPa according to Tetens' formulation (1930) and $T$ is given in degrees Celsius. Parameters $a$ and $b$ are found by empirical means. In the original formula, $a=7.5$ and $b=237.3$ \textcelsius{} over a plain of water, while $a=9.5$ and $b=265.5$ \textcelsius{} over a plain of ice. 
Although the original formula is old, it remains in use due to its simplicity and accuracy. The following equation is an improved formulation derived by \citet{bolton}:
\begin{IEEEeqnarray}{rll}
\label{eqn:B}
    e_{sat}(T) = 6.112 \exp \left(\frac{17.67 T}{T+243.5}\right)
\end{IEEEeqnarray}

showing lower errors than those obtained with the original formulation for temperatures below 0 \textcelsius{}, which is the temperature range typically found over astronomical observatories located at high-altitude sites. In the range -35 \textcelsius{} $\leq$ T $\leq$ 35 \textcelsius{} the  maximum percentage error obtained when compared to Tetens' formula is 1.68\% at -35 \textcelsius{}, which translates into a PWV percentage difference of approximately 0.025\%.
Using Eq. (\ref{eqn:B}) in Eq. (\ref{eqn:rh}), and substituting in the expression for specific humidity (Eq. \ref{eqn:shp}) one can finally compute the precipitable water vapour defined in Eq. (\ref{eqn:PWV2}) as a function of temperature, pressure and relative humidity:
\begin{IEEEeqnarray*}{rll}
    PWV &=& -\frac{1}{g \rho_{w}}\int_{p_{i}}^{p_{i+1}} \frac{0.622u \cdot 6.112 \exp \left(\frac{17.67 T}{T+243.5}\right)}{p-0.378u\cdot 6.112 \exp \left(\frac{17.67 T}{T+243.5}\right)} dp \\
    && \IEEEyesnumber
\end{IEEEeqnarray*}
In the literature one can find further approximations or simplifications to reduce the number of variables, for example by assuming an atmosphere of constant lapse rate, which occurs only in moisture free atmospheres \citep{varmaghani}.

\end{appendix}

\end{document}